\documentclass[aps,twocolumn,superscriptaddress,nofootinbib,preprintnumbers,prd,10pt]{revtex4-1}

\usepackage{mathrsfs}
\usepackage{amsfonts,amsmath}
\usepackage{array}
\usepackage{graphicx}
\usepackage{color}
\usepackage{hyperref}
\usepackage{float}

\begin{document}

\title{Probing jet medium interactions via $Z$($H$)+jet momentum imbalances}

\author{Lin Chen}
\email{chen.l.raymond@mails.ccnu.edu.cn}
\affiliation{Key Laboratory of Quark and Lepton Physics (MOE) and Institute of Particle Physics, Central China Normal University, Wuhan 430079, China}

\author{Shu-Yi Wei}
\email{shu-yi.wei@polytechnique.edu}
\affiliation{CPHT, CNRS, \'Ecole Polytechnique, Institut Polytechnique de Paris, Route de Scalay, 91128 Palaiseau, France}
\affiliation{European Centre for Theoretical Studies in Nuclear Physics and Related Areas(ECT*)\\ and Fondazione Bruno Kessler, Strada delle Tabarelle 286, I-38123 Villazzano (TN), Italy}

\author{Han-Zhong Zhang}
\email{zhanghz@mail.ccnu.edu.cn}
\affiliation{Key Laboratory of Quark and Lepton Physics (MOE) and Institute of Particle Physics, Central China Normal University, Wuhan 430079, China}

\begin{abstract}
Different types of high energy hard probes are used to extract the jet transport properties of the Quark-Gluon Plasma created in heavy-ion collisions, of which the heavy boson tagged jets are undoubtedly the most sophisticated due to its clean decay signature and production mechanism. 
In this study, we used the resummation improved pQCD approach with high order correction in the hard factor to calculate the momentum ratio $x_J$ distributions of $Z$ and Higgs($H$) tagged jets. 
We found that the formalism can provide a good description of the 5.02 TeV $pp$ data.
Using the BDMPS energy loss formalism, along with the OSU 2+1D hydro to simulate the effect of the medium, we extracted the value of the jet transport coefficient to be around $\hat{q}_0=4\sim8~GeV^2/fm$ by comparing with the $Z$+jet $PbPb$ experimental data. 
The $H$+jet $x_J$ distribution were calculated in a similar manner in contrast and found to have a stronger Sudakov effect as compared with the $Z$+jet distribution. 
This study uses a clean color-neutral boson as trigger to study the jet quenching effect and serves as a complimentary method in the extraction of the QGP's transport coefficient in high energy nuclear collisions. 
\end{abstract}

\maketitle


\section{Introduction}
\label{sec:intro}
The creation of the Quark-Gluon Plasma (QGP) is one of the most important discoveries by the Large Hadron Collider (LHC)\cite{Muller:2012zq} and the Relativistic Heavy-Ion Collider (RHIC)\cite{Gyulassy:2004zy} in recent high-energy collision experiments. Its physical properties, which exhibits almost perfect fluidity and color opaqueness, could be related to the formation and evolution of our early universe.
A large portion of the community effort is devoted to the study of the QGP's transport properties through the use of hard probes. Since energetic partonic jets will loss energy due to medium induced radiations, and gets knocked around via multiple elastic scatterings when traversing the hot and dense medium
\cite{Gyulassy:1993hr,Baier:1996kr,Baier:1996sk,Baier:1998kq,Baier:2001yt,Zakharov:1996fv,Gyulassy:1999zd,Wiedemann:2000za,Arnold:2002ja,Wang:2001ifa}, a single parameter known as the jet transport coefficient ($\hat{q}$)
\cite{Majumder:2010qh, Qin:2015srf, Blaizot:2015lma} is used to encapsulate this so-called Jet Quenching phenomena\cite{Wang:1991xy}, which is both the effect of jet energy loss, and transverse momentum broadening and is defined as the transverse momentum square transfer per unit length.
Efforts have been made to quantitatively extract this parameter that reflects the transport properties of the QGP, notably the JET collaboration by utilizing the nuclear modifications ($R_{AA}$) of single hadron yield suppressions with different energy loss models at various temperatures\cite{Burke:2013yra}. This has sparked a community wide movement in the quantitative extraction of the $\hat{q}$ variable.

It is well-known that there are two simple observables which best describe the jet quenching effect, namely the jet azimuthal angular correlation ($\Delta\phi_J=|\phi^{\rm jet}-\phi^{\rm trigger}|$) for the transverse momentum broadening effect, and the jet momentum fraction distribution ($x_J=P_\perp^{\rm jet}/P_\perp^{\rm trigger}$) for the medium induced energy loss effect. Both of which are widely used in phenomenological studies for their simplicity in calculation and measurement, and their direct relations to the transport coefficient. However, due to the nature of these two observables and its capability in describing sensitive effects, both theory and experiment have faced challenges in accurately calculating their distributions.

In the language of perturbative QCD theory\cite{Arnold:1988dp,Campbell:1999ah,Campbell:2011bn,Ravindran:2002dc,Glosser:2002gm}, both $\Delta\phi_J$ and $x_J$ distribution is expected to have a delta function at $\Delta\phi_J=\pi$ and at $x_J=1$ in the leading-order $\alpha_s$ expansion. This corresponds to the back-to-back configuration of the scattering, where the diverging behaviour is a direct consequence of the transverse momentum conservation, and it will eventually propagate to even higher orders of the perturbative series. Unfortunately, these are the regions of the distribution that is essential in the $\hat{q}$ extraction and it is imperative to employ an all order resummation to deal with these singular behaviours and setup a $pp$ baseline before using it to calculate the transport parameter.
Recent developments on the Sudakov resummation formalism\cite{Banfi:2008qs,Mueller:2012uf,Mueller:2013wwa,Sun:2014gfa,Sun:2015doa,Mueller:2016gko,Mueller:2016xoc} have demonstrated a reliable description of the experimental data in the limits of these extreme kinematic regions where pQCD would diverge. However, resummation alone is not sufficient in describing the entire regions of the phase-space especially in places where hard partonic splitting takes dominance. To overcome this challenge, one would require a clever technique to incorporate both pQCD and resummation formalisms in order to provide a better description of the experimental data.

In our previous studies\cite{Chen:2016vem,Chen:2016cof}, a resummation improved perturbative QCD approach was developed by utilizing both pQCD and resummation formalism to effectively calculate dijet momentum imbalance distribution for the numerical extraction of the transport coefficient. However, with the large error bands in our theoretical calculations, we are faced with another challenge, which is the uncertainties that comes when both outgoing jets gets quenched. The cross-section does not discriminate the species of the jets, where in fact quark and gluon jets quench differently by their color factor, and we can only assume that all jets are either quarks or gluons in our calculation. To encounter this, we fixed one of the outgoing particle by using the gamma-jet correlation\cite{Chen:2018fqu}, whereby the color neutral photon does not participate in any medium interactions, and thus the simple cross-section allows us to implement different quenching factors for individual quarks and gluons species.

Even so, experimental uncertainties showed that photons suffer heavy contaminations from sources such as initial state, fragmentation and thermal radiations which makes it very difficult to isolate the photon that is coming from the actual hard scattering. Furthermore, experimental measurements suffer from detector effects that causes bin migration in the $x_J$ distribution and other $P_\perp(E_\perp)$ sensitive observables, which needs to be taken into account before a direct comparison between theoretical calculation and experimental measurements can take place.

In this study, we use heavy neutral boson $B(Z,H)$ as trigger, correlating with an associate jet as probe to extract the transport coefficient $\hat{q}$, with main focus on $Z$+jet correlation. Similar to the photon-jet correlations, the weak $Z$ boson does not interact strongly with the QCD medium, and with its life time much longer than that of the QGP produced in current accelerators, the $Z$ boson preserves the momentum informations of the away-side jets before getting quenched. In contrast, it is produced almost entirely from hard scatterings due to its heavy mass, and with its clean dileptonic decay signatures, it could be regarded as the standard candle in high energy collisions.
It is worth mentioning that the current study is complementary to those based on the use of Monte-Carlo event generators
\cite{Zhang:2018urd,Casalderrey-Solana:2015vaa,KunnawalkamElayavalli:2016ttl,Kang:2017xnc}, which matches hard matrix-element to parton showers that mimics the effect of multiple soft radiations. In the limit of infinite branching, parton shower should be equivalent to the framework of resummation.
Although its production yield is suppress by its heavy mass, the era of the LHC has provided rich statistics for us to utilize this golden probe in the study of heavy-ion physics.


\section{Resummation formalism}
\label{sec:form}
We begin by stating that the genesis of the singularity aforementioned occurs from the scale hierarchy of $Q^2\gg q_\perp^2$ in which $Q^2$ is the hard scale and $q_\perp$ the transverse momentum imbalance of the scattering system defined as $\vec{q}_\perp\equiv \vec{P}_{B\perp}+\vec{P}_{J\perp}$. A typical configuration puts $Q^2$ of the order of the jet momentum, and $q_\perp$ the overall transverse momentum kick of the soft radiations. At the LHC, the above configuration can have large logarithmic terms in the form of $\alpha_s^n\ln^{2n-1}(Q^2/q_\perp^2)$ known as the Sudakov double logarithms, which will appear in every order of the conventional perturbative QCD calculations in the $\alpha_s$ expansion. Thus, critical phase space regions like $\Delta\phi_J$ distributions near $\pi$, or back-to-back $x_{J}$ distributions near 1 would fail to converge. This calls for a resummation technique that could effectively resum multiple vacuum soft gluon emissions which contribute to the overall $q_\perp$ kick.

The $q_\perp$ resummation technique was originally developed in the Drell-Yan framework for heavy boson production\cite{Collins:1984kg,Yuan:1991we,Qiu:2000ga}, and recent studies have extended it to include jets in the final state. This includes Dijet production\cite{Sun:2014gfa,Sun:2015doa}, Higgs+jet production\cite{Sun:2014lna,Sun:2016kkh}, and $Z$+jet production\cite{Sun:2018icb}. In this study, both $Z$+jet and $H$+jet resummation up to 1-loop order will be used for analysis. The multi-differential all-order resummation cross-section for the $p+p\rightarrow B(P_B)+{\rm Jet}(P_J)+X$ process is given as\cite{Sun:2018icb}:
\begin{eqnarray}\label{eq:master}
&&\frac{d^5\sigma}{dy_Bdy_JdP_{J\perp}^2d^2\vec{q}_\perp}=\sum_{ab}\sigma_0\nonumber\\
&&\left[\int\frac{d^2\vec{b}_\perp}{(2\pi)^2}e^{-i\vec{q}_\perp\cdot\vec{b}_\perp}W_{ab\rightarrow BJ}(x_1,x_2,b_\perp)+Y_{ab\rightarrow BJ}\right]
\end{eqnarray}
with
\begin{eqnarray}
W_{ab\rightarrow BJ}
&=&x_1f_a(x_1,\mu_{\rm fac})~x_2f_b(x_2,\mu_{\rm fac})\nonumber\\
&\times& e^{-\mathcal{F}_{\rm NP}}e^{-S_{\rm Sud}(s,\mu_{\rm res})}H_{ab\rightarrow BJ}(s,\mu_{\rm res})~.
\end{eqnarray} 

Here $y_B$ and $y_J$ are the rapidities of the heavy boson and jet respectively. $P_{B\perp}$ and $P_{J\perp}$ are the boson and jet's transverse momentum. $\sigma_0$ is the normalization factor for the particular $pp\rightarrow B+jet$ process. 
The auxiliary $b_\perp$-space integral guarantees transverse momentum conservation of the radiated gluons. 
With $W$-term the all order resummation term and $Y$-term the fixed order correction term. In this study, we will neglect the contribution of the $Y$-term with reasons to be explain later. In the $W$-term, $x_{1,2}=(Q_Be^{\pm y_B}+Q_Je^{\pm y_J})/\sqrt{S}$ denotes the momentum fraction of the incoming parton from its parent hadron, with $Q_B^2=m_B^2+P_{B\perp}^2$ and $Q_J^2=P_{J\perp}^2$ the boson and jet transverse scale respectively, while $\sqrt{S}$ denotes the usual collision energy in the Center-of-Mass frame. $f_{a,b}$ are the parton distribution functions (PDFs) of the incoming parton species $a$ and $b$. The factorized hard part is represented through the hard factor ($H$), while the soft part is captured by the Sudakov factor ($S_{\rm Sud}$).

The 1-loop order standard Sudakov form factor is expressed as follows\cite{Sun:2018icb}:
\begin{eqnarray}
&&S_{\rm Sud}=\int_{\mu_{\rm fac}^2}^{\mu_{\rm res}^2}\frac{d\mu^2}{\mu^2}\nonumber\\
&&\left[\left(A^{(1)}+A^{(2)}\right)\ln\frac{s}{\mu^2}+B_1^{(1)}+B_2^{(1)}+D^{(1)}\ln\frac{1}{R^2}\right]
\end{eqnarray}
where the integral solves the energy evolution of the soft factor from the factorization scale $\mu_{\rm fac}$ to the resummation scale $\mu_{\rm res}$. The $A$ and $B_1$ terms reflects the color exchanges between the incoming partons, thus will depend only on the incoming parton species. An additional $B_2$ term is included in contrast to Drell-Yan processes that reflects the color interactions between the incoming partons and the outgoing quark jet. The $D$ term takes care of the soft radiations outside of the jet with cone size $R$. We summarize these perturbatively calculable terms in Table \ref{pert}, with coefficients $K=\left(\frac{67}{18}-\frac{\pi^2}{6}\right)C_A-\frac{10}{9}N_fT_R$ and $\beta_0=\frac{11-2/3N_f}{12}$ found in the reference\cite{deFlorian:2000pr,Catani:2000vq,deFlorian:2001zd}.
\begin{table}
\begin{center}
\begin{tabular}{|c|c|c|}\hline
			&	quark						&	gluon							\\\hline
$A^{(1)}$	&	$C_F\left(\frac{\alpha_s}{2\pi}\right)$					&	$C_A\left(\frac{\alpha_s}{2\pi}\right)$						\\\hline
$A^{(2)}$	&	$K~C_F\left(\frac{\alpha_s}{2\pi}\right)^2$				&	$K~C_A\left(\frac{\alpha_s}{2\pi}\right)^2$					\\\hline
$B_1^{(1)}$	&	$-\frac{3}{2}C_F\left(\frac{\alpha_s}{2\pi}\right)$		&	$-2\beta_0C_A\left(\frac{\alpha_s}{2\pi}\right)$				\\\hline
$B_2^{(1)}$	&	$\ln\frac{u}{t}C_F\left(\frac{\alpha_s}{2\pi}\right)$	&	$-\ln\frac{u}{t}C_A\left(\frac{\alpha_s}{2\pi}\right)$		\\\hline
$D^{(1)}$	&	$C_F\left(\frac{\alpha_s}{2\pi}\right)$					&	$C_A\left(\frac{\alpha_s}{2\pi}\right)$						\\\hline
\end{tabular}
\end{center}
\caption{Perturbative coefficients for the Sudakov integral, where one sums the $A$ and $B$ factor for incoming parton species, and choose $D$ for the corresponding jet species.}
\vspace*{-5mm}
\label{pert}
\end{table}

The Mandelstam variables are defined in the usual manner, and can be simplify to the following for the above processes:
\begin{eqnarray}
s&=&(P_a+P_b)^2=x_1x_2S~,\\
t&=&(P_a-P_B)^2=m_B^2-x_1\sqrt{S}Q_Be^{-y_B}~,\\
u&=&(P_a-P_J)^2=-x_1\sqrt{S}Q_Je^{-y_J}~.
\end{eqnarray}
One must sum the corresponding $A$ and $B$ terms for both incoming parton species and choose $D$ for the corresponding jet species. 
Note that the strong coupling $\alpha_s(\mu)$ runs in the above $d\mu$ integral. In our numerical calculation, the Sudakov integral is solved exactly with the 2-loop running coupling which includes both $b_0$ and $b_1$ terms.

To prevent the $b_\perp$ integral from hitting the non-perturbative region $q_\perp^2\lesssim \Lambda_{\rm QCD}^2$, the $b_*=b/\sqrt{1+b^2/b_{\rm max}^2}$ prescription is introduced
\cite{Davies:1984sp,Ladinsky:1993zn,Landry:2002ix,Su:2014wpa,Prokudin:2015ysa} , with $b_{\rm max}=1.5~GeV^{-1}$. The result of this cut-off is the definition of the factorisation scale $\mu_{\rm fac}=b_0/b_*$ with $b_0=2e^{-\gamma_E}$, and an additional non-perturbative exponent in the form:
\begin{equation}
\mathcal{F}_{\rm NP}(Q^2,b_*)=g_1b^2+g_2\ln\frac{\mu_{\rm res}}{Q_0}\ln\frac{b}{b_*}
\end{equation}
where $g_1=0.212$, $g_2=0.84$ and $Q_0^2=2.4~GeV^2$ are values fitted phenomenologically in the reference\cite{Su:2014wpa}.

We first consider the two leading-order diagrams of the $Z$+jet production in Fig. \ref{fig:Zjet} below:
\vspace*{1mm}
\begin{figure}[H]
\begin{center}
\includegraphics[width=0.39\linewidth]{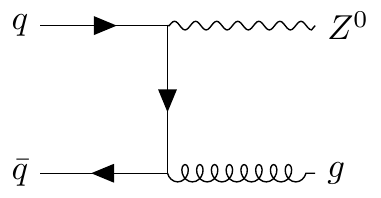}
\hfill
\includegraphics[width=0.39\linewidth]{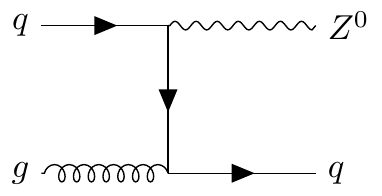}
\end{center}
\caption{Two examples of the lowest order partonic scattering diagram for $Z$+jet production.}
\label{fig:Zjet}
\end{figure}
\vspace*{-5mm}
\noindent The normalization factor for this process is\cite{Sun:2018icb}:
\begin{equation}
\sigma_0^{Z+\rm jet}=\frac{\alpha_s(\mu_{\rm ren})(g_V^2+g_A^2)}{16s^2}
\end{equation}
where the $Z$ to quark coupling is represented through the vector and axial-vector gauge couplings\cite{Ellis:1991qj}:
\begin{equation}
g_V=\frac{g_W}{2\cos\theta_W}(\tau_3^q-2Q_q\sin^2\theta_W),~~~g_A=\frac{g_W}{2\cos\theta_W}\tau_3^q
\end{equation}
with $g_W=\sqrt{\frac{\alpha_e4\pi}{\sin^2\theta_W}}$ the weak gauge coupling and $\theta_W$ the weak Weinberg angle. Here, $\cos^2\theta_W=m_W^2/m_Z^2$, $Q_q$ is the quark electric charge and $\tau_3^q$ the third component of the quark weak isospin.

The hard factor can be expanded in a perturbative series:
\begin{eqnarray}
H_{ab\rightarrow BJ}
&=&H^{(0)}_{ab\rightarrow BJ}+H^{(1)}_{ab\rightarrow BJ}+\cdots\nonumber\\
&=&H^{(0)}_{ab\rightarrow BJ}\left(1+\frac{\alpha_s(\mu_{\rm ren})}{2\pi}[\cdots]+\cdots\right)~,
\end{eqnarray}
for the $q\bar{q}\rightarrow Zg$ channel, the leading and one-loop order hard factors are as follows\cite{Sun:2018icb}:
\begin{equation}
H_{q\bar{q}\rightarrow Zg}^{(0)}=\frac{8}{3}C_F\left[\frac{t^2+u^2+2m_Z^2s}{tu}\right]~,
\end{equation}
\begin{widetext}
\begin{eqnarray}
H_{q\bar{q}\rightarrow Zg}^{(1)}
&=&H_{q\bar{q}\rightarrow Zg}^{(0)}\frac{\alpha_s}{2\pi}\left\{\left[-2\beta_0\ln\left(\frac{R^2P_{J\perp}^2}{\mu_{\rm res}^2}\right)+\frac{1}{2}\ln^2\left(\frac{R^2P_{J\perp}^2}{\mu_{\rm res}^2}\right)+{\rm Li}_2\left(\frac{m_Z^2}{m_Z^2-t}\right)+{\rm Li}_2\left(\frac{m_Z^2}{m_Z^2-u}\right)\right.\right.\nonumber\\
&&-\ln\left(\frac{\mu_{\rm res}^2}{m_Z^2}\right)\ln\left(\frac{sm_Z^2}{tu}\right)-\frac{1}{2}\ln^2\left(\frac{\mu_{\rm res}^2}{m_Z^2}\right)+\frac{1}{2}\ln^2\left(\frac{s}{m_Z^2}\right)-\frac{1}{2}\ln^2\left(\frac{tu}{m_Z^4}\right)+\ln\left(\frac{-t}{m_Z^2}\right)\ln\left(\frac{-u}{m_Z^2}\right)\nonumber\\
&&\left.+\frac{1}{2}\ln^2\left(\frac{m_Z^2-t}{m_Z^2}\right)+\frac{1}{2}\ln^2\left(\frac{m_Z^2-u}{m_Z^2}\right)-\frac{1}{2}\ln^2\left(\frac{1}{R^2}\right)-\frac{2\pi^2}{3}+\frac{67}{9}-\frac{23N_f}{54}\right]C_A+6\beta_0\ln\frac{\mu_{\rm ren}^2}{\mu_{\rm res}^2}\nonumber\\
&&\left.+\left[2\ln\left(\frac{s}{m_Z^2}\right)\ln\left(\frac{\mu_{\rm res}^2}{m_Z^2}\right)-\ln^2\left(\frac{\mu_{\rm res}^2}{m_Z^2}\right)-3\ln\left(\frac{\mu_{\rm res}^2}{m_Z^2}\right)-\ln^2\left(\frac{s}{m_Z^2}\right)+\pi^2-8\right]C_F\right\}~.
\end{eqnarray}
\end{widetext}
In the above hard factors, ${\rm Li}_2(z)$ denotes the dilogarithm (Spence's) function, and $\mu_{\rm ren}$ is the renormalization factor.
Note that the gluon($C_A$) term is dependent on the jet parameter indicating a final state gluon jet. Similarly for the $qg\rightarrow Zq$ channel, we have the corresponding leading and one-loop hard factors \cite{Sun:2018icb}:\\
\begin{equation}
H_{qg\rightarrow Zq}^{(0)}=C_F\left[\frac{s^2+t^2+2m_Z^2u}{-st}\right]~,
\end{equation}
\begin{widetext}
\begin{eqnarray}
H_{qg\rightarrow Zq}^{(1)}
&=&H_{qg\rightarrow Zq}^{(0)}\frac{\alpha_s}{2\pi}\left\{\left[-{\rm Li}_2\left(\frac{m_Z^2}{s}\right)+{\rm Li}_2\left(\frac{m_Z^2}{m_Z^2-t}\right)-\ln\left(\frac{\mu_{\rm res}^2}{m_Z^2}\right)\ln\left(\frac{um_Z^2}{st}\right)-\frac{1}{2}\ln^2\left(\frac{\mu_{\rm res}^2}{m_Z^2}\right)\right.\right.\nonumber\\
&&-\frac{1}{2}\ln^2\left(\frac{-st}{m_Z^4}\right)+\ln\left(\frac{s}{m_Z^2}\right)\ln\left(\frac{s-m_Z^2}{m_Z^2}\right)-\frac{1}{2}\ln^2\left(\frac{s}{m_Z^2}\right)+\frac{1}{2}\ln^2\left(\frac{m_Z^2-t}{m_Z^2}\right)+\frac{1}{2}\ln^2\left(\frac{-u}{m_Z^2}\right)\nonumber\\
&&\left.+\frac{\pi^2}{2}\right]C_A+6\beta_0\ln\frac{\mu_{\rm ren}^2}{\mu_{\rm res}^2}+\left[-\frac{3}{2}\ln\left(\frac{R^2P_{J\perp}^2}{\mu_{\rm res}^2}\right)+\frac{1}{2}\ln^2\left(\frac{R^2P_{J\perp}^2}{\mu_{\rm res}^2}\right)+2\ln\left(\frac{-u}{m_Z^2}\right)\ln\left(\frac{\mu_{\rm res}^2}{m_Z^2}\right)\right.\nonumber\\
&&\left.\left.-\ln^2\left(\frac{\mu_{\rm res}^2}{m_Z^2}\right)-3\ln\left(\frac{\mu_{\rm res}^2}{m_Z^2}\right)-\ln^2\left(\frac{-u}{m_Z^2}\right)-\frac{1}{2}\ln^2\left(\frac{1}{R^2}\right)-\frac{2\pi^2}{3}-\frac{3}{2}\right]C_F\right\}~.
\end{eqnarray}
\end{widetext}

We then consider the leading dominant channels of $H$+jet production in Fig. \ref{fig:Hjet} below:
\begin{figure}[H]
\begin{center}
\includegraphics[width=0.44\linewidth]{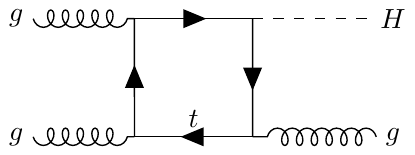}
\hfill
\includegraphics[width=0.44\linewidth]{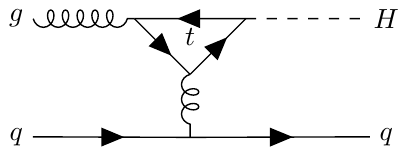}
\end{center}
\caption{Two examples of the lowest order partonic scattering diagram for $H$+jet production.}
\label{fig:Hjet}
\end{figure}
\noindent The normalization factor for this process is\cite{Sun:2016kkh}:
\begin{equation}
\sigma_0^{H+\rm jet}=\frac{4}{9}\frac{4\alpha_s^3(\mu_{\rm ren})\sqrt{2}G_F}{s^2(4\pi)^3}
\end{equation}
where $G_F$ is the Fermi constant. The Higgs are produced through the top quark loop, and the heavy top mass limit is applied to describe the effective coupling between quarks and the SM Higgs. The leading order hard factor is expressed as\cite{Sun:2016kkh}:
\begin{eqnarray}
H_{gg\rightarrow Hg}^{(0)}&=&\frac{C_A}{4(N_c^2-1)}\left[\frac{s^4+t^4+u^4+m_H^8}{stu}\right]~,\\
H_{qg\rightarrow Hq}^{(0)}&=&\frac{C_F}{4(N_c^2-1)}\left[\frac{s^2+t^2}{-u}\right]~,
\end{eqnarray}
with the one-loop order hard factor as follows:
\begin{widetext}
\begin{eqnarray}
H_{gg\rightarrow Hg}^{(1)}
&=&H_{gg\rightarrow Hg}^{(0)}\frac{\alpha_s}{2\pi}\left[-2\beta_0\ln\left(\frac{R^2P_{J\perp}^2}{\mu_{\rm res}^2}\right)+\ln^2\left(\frac{\mu_{\rm res}^2}{P_{J\perp}^2}\right)+\ln\left(\frac{1}{R^2}\right)\ln\left(\frac{\mu_{\rm res}^2}{P_{J\perp}^2}\right)+6\beta_0\ln\frac{\mu_{\rm ren}^2}{\mu_{\rm res}^2}-2\ln\left(\frac{P_{J\perp}^2}{\mu_{\rm res}^2}\right)\ln\left(\frac{s}{\mu_{\rm res}^2}\right)\right.\nonumber\\
&&-2\ln\left(\frac{s}{-t}\right)\ln\left(\frac{s}{-u}\right)+\ln^2\left(\frac{m_H^2-t}{m_H^2}\right)-\ln^2\left(\frac{m_H^2-t}{-t}\right)+\ln^2\left(\frac{m_H^2-u}{m_H^2}\right)-\ln^2\left(\frac{m_H^2-u}{-u}\right)\nonumber\\
&&\left.+2{\rm Li}_2\left(1-\frac{m_H^2}{s}\right)+2{\rm Li}_2\left(\frac{t}{m_H^2}\right)+2{\rm Li}_2\left(\frac{u}{m_H^2}\right)+\frac{67}{9}+\frac{\pi^2}{2}-\frac{23N_f}{54}\right]C_A~,\\
H_{qg\rightarrow Hq}^{(1)}
&=&H_{qg\rightarrow Hq}^{(0)}\frac{\alpha_s}{2\pi}\left\{\left[\frac{1}{2}\ln^2\left(\frac{\mu_{\rm res}^2}{P_{J\perp}^2}\right)+\ln\left(\frac{P_{J\perp}^2}{\mu_{\rm res}^2}\right)\ln\left(\frac{u}{t}\right)+\ln\left(\frac{P_{J\perp}^2}{\mu_{\rm res}^2}\right)\ln\left(\frac{s}{\mu_{\rm res}^2}\right)-2\ln\left(\frac{-t}{\mu_{\rm res}^2}\right)\ln\left(\frac{-u}{\mu_{\rm res}^2}\right)\right.\right.\nonumber\\
&&\left.-4\beta_0\ln\left(\frac{-u}{\mu_{\rm res}^2}\right)+6\beta_0\ln\frac{\mu_{\rm ren}^2}{\mu_{\rm res}^2}+2{\rm Li}_2\left(\frac{u}{m_H^2}\right)-\ln^2\left(\frac{m_H^2-u}{-u}\right)+\ln^2\left(\frac{m_H^2-u}{m_H^2}\right)+\frac{7+4\pi^2}{3}\right]C_A\nonumber\\
&&+20\beta_0+\left[\frac{1}{2}\ln^2\left(\frac{\mu_{\rm res}^2}{P_{J\perp}^2}\right)+\frac{3}{2}\ln\left(\frac{\mu_{\rm res}^2}{R^2P_{J\perp}^2}\right)+\ln\left(\frac{1}{R^2}\right)\ln\left(\frac{\mu_{\rm res}^2}{P_{J\perp}^2}\right)-\ln\left(\frac{P_{J\perp}^2}{\mu_{\rm res}^2}\right)\ln\left(\frac{u}{t}\right)\right.\nonumber\\
&&-\ln\left(\frac{P_{J\perp}^2}{\mu_{\rm res}^2}\right)\ln\left(\frac{s}{\mu_{\rm res}^2}\right)+3\ln\left(\frac{-u}{\mu_{\rm res}^2}\right)+2{\rm Li}_2\left(1-\frac{m_H^2}{s}\right)+2{\rm Li}_2\left(\frac{t}{m_H^2}\right)-\ln^2\left(\frac{m_H^2-t}{-t}\right)\nonumber\\
&&\left.\left.+\ln^2\left(\frac{m_H^2-t}{m_H^2}\right)-\frac{3}{2}-\frac{5\pi^2}{6}\right]C_F\right\}~.
\end{eqnarray}
\end{widetext}

\begin{figure}[H]
\begin{center}
\includegraphics[width=0.8\linewidth]{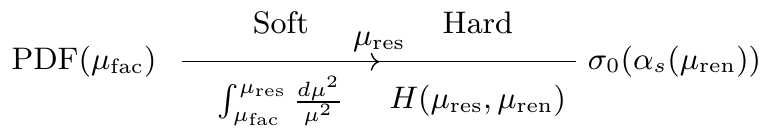}
\end{center}
\caption{graphical depiction of hierarchical structure between the different scales.}
\label{fig:scale}
\end{figure}
We note in Fig. \ref{fig:scale} that three distinct scales are involved in this formalism.
The factorization scale $\mu_{\rm fac}$ appearing in the PDF is fixed at $\mu_{\rm fac}=b_0/b_*$.
The renormalization scale $\mu_{\rm ren}$ that appears in $\sigma_0$ is taken to be $\mu_{\rm ren}=H_T=Q_B+Q_J$. Since a reliable theory should be insensitive to the choice of the renormalization scale, we varied this by a factor of $2^{\pm1}$ and found little difference numerically.
There is a freedom of choice for the resummation scale that varies between the two fixed scales to control the evolution of the soft and hard part simultaneously.
However, the choice of the resummation scale is not trivial, since it enters the hard factor via double logarithmic terms($\ln^2(\mu_{\rm res}/P_{J\perp}^2)$). In order to minimise the contributions from these possible large logs, we will set the resummation scale to $\mu_{\rm res}=P_{J\perp}$\cite{Sun:2016kkh}.

In the calculation above, the familiar color factors are used:
\begin{equation}
N_c=3;~C_A=N_c;~C_F=\frac{N_c^2-1}{2N_c};~T_R=\frac{1}{2}
\end{equation}
with the values of the boson masses($m_Z,m_W,m_H$) and the Fermi-coupling constant($G_F$) taken from PDG\cite{Tanabashi:2018oca}.


\section{vacuum and smearing}
\label{sec:pp}

Assuming the absence of medium effects in hadron-hadron collisions, we now have an all-order resummed calculation up to one-loop order that is best at describing near back-to-back events in $pp$ collisions corresponding to data near $\pi$ in the $\Delta\phi_J$ distributions, and good descriptions of several experimental data were achieved at various collision energies ranging from 1.8 TeV\cite{Aaltonen:2014vma,Abazov:2009pp} to 8 TeV\cite{Chatrchyan:2013tna,Khachatryan:2016crw}.
However, our previous investigations have shown that the $\Delta\phi_J$ distribution is great for observing the medium induced broadening effect only at lower kinematic regimes such as RHIC, due to the fact that medium broadening effects were dwarfed by the overwhelming vacuum Sudakov broadening at the LHC energy. This shifts our attention to the more sophisticated $x_J$ variable.

\begin{figure}
\begin{center}
\includegraphics[width=0.8\linewidth]{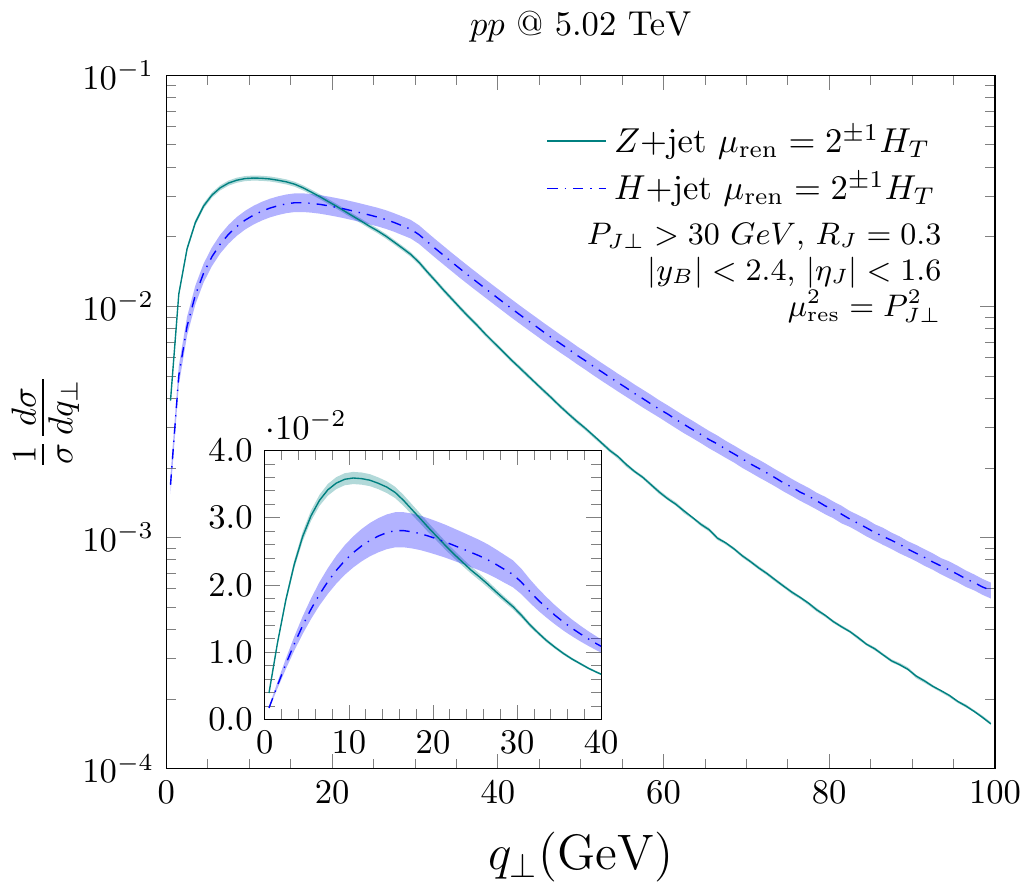}
\end{center}
\caption{Normalized $q_\perp$ distribution for both $Z$+jet(solid) and $H$+jet(dashdotted) processes at 5.02 TeV. The resummation scale is fixed at $\mu_{\rm res}^2=P_{J\perp}^2$, while the renormalization scale varies by a factor $\mu_{\rm ren}^2=2^{\pm1}H_T^2$.}
\label{fig:502TZHqt}
\end{figure}
We begin first with the analysis on the strength of the Sudakov effect by plotting the $q_\perp$ distribution for both $Z$ and Higgs plus a jet production shown in Fig. \ref{fig:502TZHqt}. One would expect perturbative calculations to diverge as $q_\perp$ approaches 0 due to transverse momentum conservation. We see that the Higgs+jet correlation indicated by the dashdotted line has a broader spectrum than that of $Z$+jet correlations indicated by the solid line. We know that Higgs has a higher mass and is dominated by $gg$ channel, where $Z$+jet production is dominated by $qg$ channel, and as a result, Higgs is shown to have a stronger Sudakov effect than $Z$. This tells us that Higgs has a higher tendency of radiating soft gluons that contributes to higher overall $q_\perp$ than $Z$. By varying the renormalization scale with a factor of $2^{\pm1}$, we see that $Z$ is less scale sensitive than $H$ indicating that higher order calculations for the $H$+jet process is needed for precision measurements.

Unlike the $\Delta\phi_J$ distribution where we can approximate sections of the spectrum near $\pi$ to be dominated by back-to-back processes, the spectrum of the $x_J$ distribution is superpositioned by different processes that one would require both resummation and perturbative calculation in order to have a good description of the entire region of the distribution. And the method that we employed is the so-called resummation improved pQCD approach where the switching between the two formalisms is determined by a $\phi_m$ cut on $\Delta\phi_J$, in which the position of $\phi_m$ is the intersection of the two calculations. This is demonstrated in the following equation:
\begin{eqnarray}
\frac{1}{\sigma}\frac{d\sigma_{\rm improved}}{dx_J}
&=&\frac{1}{N}\left(\left.\frac{1}{\sigma_{\rm pQCD}}\frac{d\sigma_{\rm pQCD}}{dx_J}\right|_{\Delta\phi<\phi_m}\right.\nonumber\\
&&\left.+\left.\frac{1}{\sigma_{\rm res}}\frac{d\sigma_{\rm res}}{dx_J}\right|_{\phi_m<\Delta\phi<\pi}\right)
\end{eqnarray}

Note that in the language of pQCD, leading-order total cross-section which evaluates 2-to-2 subprocesses, corresponds to the trivial order in $\Delta\phi$ differential cross-section. Thus the next-to-leading-order calculations which evaluates 2-to-3 subprocesses corresponds to $\Delta\phi$ distributions at leading-order(LO) shown in the plots.

\begin{figure}
\begin{center}
\includegraphics[width=0.8\linewidth]{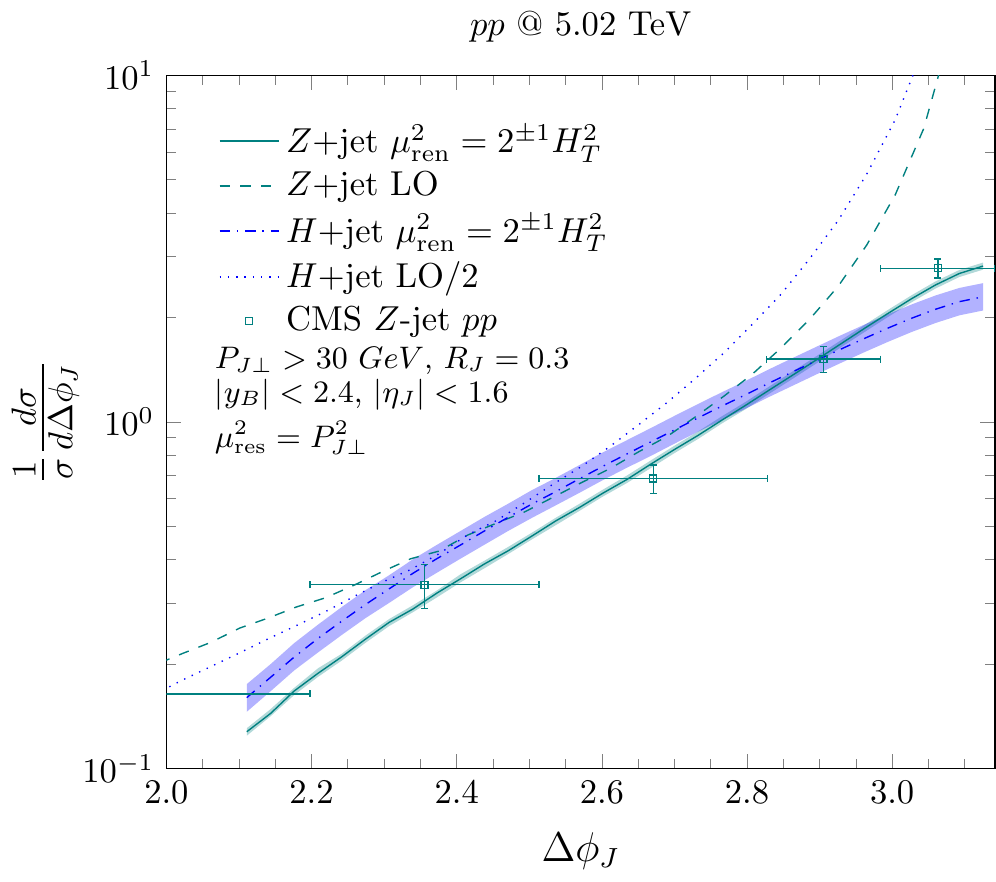}
\end{center}
\caption{Normalized azimuthal angular $\Delta\phi$ distribution using resummation and perturbative calculation for $Z$+jet process(solid and dashed) and $H$+jet process(dashdotted and dotted) at 5.02 TeV. Data from the CMS\cite{Sirunyan:2017jic} experiment is shown in comparison.}
\label{fig:5020ppdp}
\end{figure}
We then plotted the resummed and perturbative calculations for the azimuthal distribution, both $Z$(solid and dashed line) and $H$(dashdotted and dotted line) plus a jet production shown in Fig. \ref{fig:5020ppdp} at 5.02 TeV in comparison with the available experimental data from CMS\cite{Sirunyan:2017jic}. Clearly the perturbative calculation will diverge near $\Delta\phi_{J}\approx\pi$ and we see that the resummed calculation has a better description of the experimental data in this region. Also, by merging the LO perturbative calculations with the resummed results, the choice of the $\phi_m$ switch can be placed in the vicinity of $7\pi/8$ for $Z$+jet, and $6\pi/8$ for $H$+jet. This also the evidence that soft radiations from Higgs+jet processes with a stronger Sudakov effects will lead to a broader spectrum and a larger region of the back-to-back azimuthal distribution not being able to describe by perturbative calculation.

\begin{figure}
\begin{center}
\includegraphics[width=0.84\linewidth]{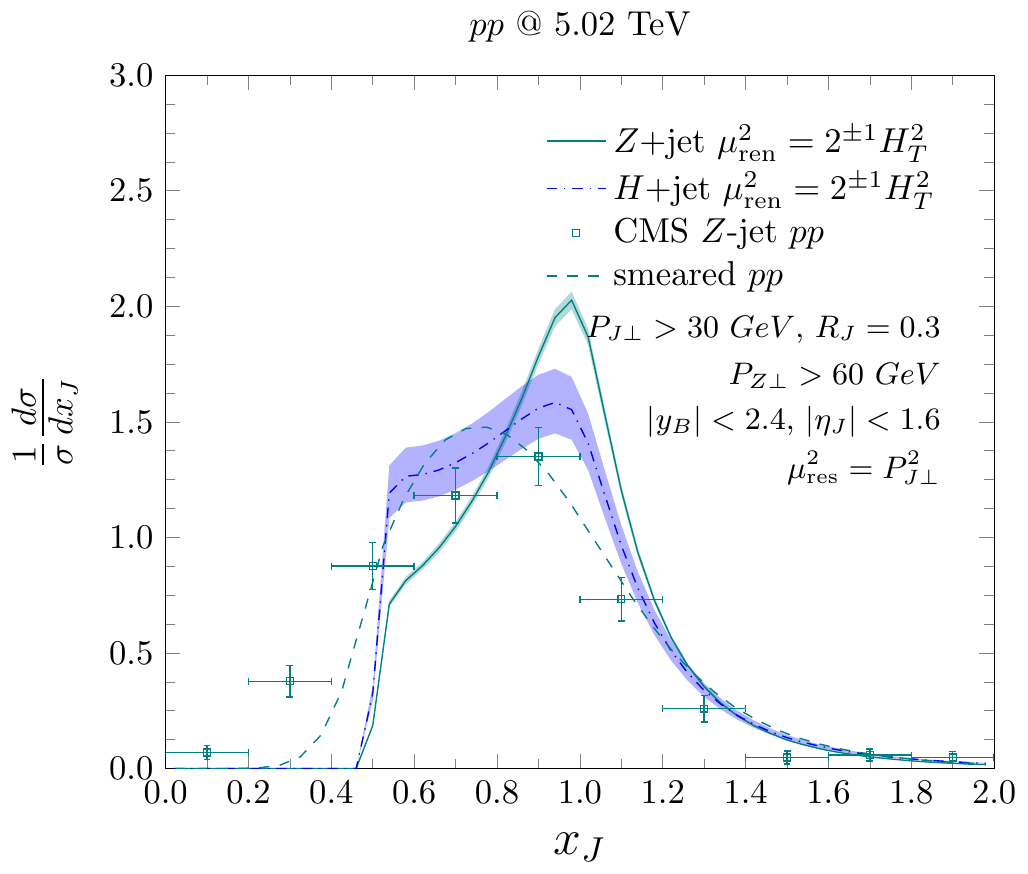}
\end{center}
\caption{Normalized momentum ratio $x_{JZ}$ distribution calculation for both $Z$+jet process(solid) and $H$+jet process(dashdotted) at 5.02 TeV. Data from the CMS\cite{Sirunyan:2017jic} experiment is shown in comparison with the smeared(dahsed) $Z$+jet result.}
\label{fig:5020ppxj}
\end{figure}

To illustrate that the resummation cross-section gives dominating yield for the $x_{JZ}$ distribution, we plotted in Fig. \ref{fig:5020ppxj} the resummed results for $Z$+jet(solid line) and $H$+jet(dashdotted) along with the smeared distribution for $Z$+jet(dashed line). As mentioned before, Higgs with stronger soft radiations will result larger imbalance in the momentum ratio than $Z$, thus the broader spectrum. We note that the $x_J$ distribution of $\gamma$+jet correlations\cite{Chen:2018fqu} has a clear small $x_J$ shoulder due to the contribution of higher order hard splittings which must be evaluate using perturbative calculations. However in this plot, only a back-to-back peak with a Sudakov tail is visible. This tells us that events in this region($7\pi/8<\Delta\phi<\pi$) are mostly back-to-back and we can approximate their distribution with resummation, which is also the reason why we have dropped the contribution of the $Y$-term in Eq. \ref{eq:master}.

As mentioned before in our previous studies, a smearing function is to be introduced to convolute with our calculation. The reason for this is that, unlike the $\Delta\phi$ distribution which measures only the position of the outgoing particles, the $x_{JZ}$ distribution measures the energy deposits of the outgoing jet of hadrons and a pair of dilepton decays from $Z$. The mis-measurement of the transverse energy deposited in the calorimeter can cause momentum bin migration of the measured particles. This poses a great challenge for the detector itself and the response effect is known to smear out sharp symmetric distributions. In this study, we use a Gaussian form smearing function with mean $\bar{r}=0.92$ and width $\sigma=0.20$ to smear our final cross-section as shown below:
\begin{eqnarray}
\frac{d\sigma_{\rm smeared}}{dP_{J\perp}}=\int \frac{dr}{\sqrt{2\pi}\sigma}e^{-\frac{(r-\bar{r})^2}{2\sigma^2}}\frac{1}{r}\left.\frac{d\sigma}{dP_{J\perp}'}\right|_{P_{J\perp}=rP_{J\perp}'}
\end{eqnarray}

We remind readers that a simple Gaussian smearing function only approximates the effect of detector response, and serves only as a comparative reference between theoretical calculation and experimental measurement. Since this response is difficult to disentangle, a proper way to compare between theory and experiment results is by using corrected data through the so-called unfolding process. This can be done when future unfolded data is available, and the current theoretical calculation will serve as a benchmark for comparison.

Note that there is a shift of the distribution peak towards small $x_{JZ}$, besides from the effect of the smearing function that we have implemented, the shift is also caused by the asymmetry in the transverse momentum cuts of the boson and the jet that rejected some events with large $x_{J\perp}$.


\section{medium and quenching}
\label{sec:aa}

We now have a solid theoretical framework for calculating differential cross-sections of heavy boson tagged jets in $pp$ collisions, and we have also fixed our smearing parameter by comparing with existing experimental data. We can now include the effect of the QGP medium by employing the BDMPS\cite{Baier:1996kr,Baier:1996sk,Baier:1998kq,Baier:2001yt} energy-loss formalism represented as follows:
\begin{equation}
\epsilon D(\epsilon)=\sqrt{\frac{\alpha^2\omega_c}{2\epsilon}}\exp\left[-\frac{\pi\alpha^2\omega_c}{2\epsilon}\right]\label{eq:radprob}
\end{equation}
where $D(\epsilon)$ is the radiation probability as a function of the radiated energy $\epsilon$. With $\alpha\equiv\frac{2\alpha_s(\mu_{\rm ren}^2)C_R}{\pi}$ for quark ($C_R=C_F$) and gluon($C_R=C_A$) jets. The characteristic gluon radiation frequency($\omega_c$) is related to the transport coefficient through the following:
\begin{equation}
\omega_c(x,y,\psi)=\int\hat{q}_R(\tau)\tau d\tau,~\hat{q}_q=\hat{q}_0\frac{T^3}{T_0^3},~\hat{q}_g=\hat{q}_q\frac{C_A}{C_F}
\end{equation}
where $\hat{q}_0$ is the quark jet transport coefficient at the center of the fireball at proper time $\tau=\tau_0$, and $T_0=T(0,0,\tau_0)$. Here we assume a simple temperature scaling of the transport coefficient $\hat{q}_q/T^3=\hat{q}_0/T_0^3$, and $\hat{q}_R=\hat{q}_q,~\hat{q}_g$ is the quenching parameter of the jet with the corresponding species.

\begin{figure}
\begin{center}
\includegraphics[width=0.8\linewidth]{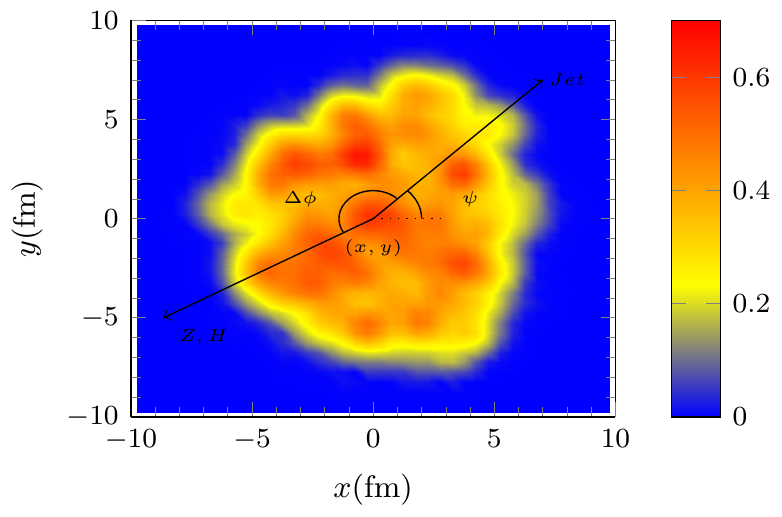}
\end{center}
\caption{graphical depiction of a $Z$ or $H$ trigger jet process going in an almost back-to-back configuration by an angle $\Delta\phi$ in the transverse plane. The position and orientation of this hard scattering in the medium is governed by its coordinate of origin $(x,y)$ and azimuthal angle $\psi$ of the jet. The colorbar represents the local temperature in GeV.}
\label{fig:profile}
\end{figure}
As in our previous studies, we employed OSU 2+1D VISH code\cite{Song:2007ux,Qiu:2011hf} to simulate the space-time evolution of the medium and, using its temperature profile, generate the radiation frequency $\omega_c$ profile as a function of transverse position of the scattering and orientation of the jet as depicted in Fig. \ref{fig:profile}. Substituting this radiation probability with the corresponding color factors for the different species of the outgoing jet, a final integration over the geometry is performed to give the quenched result:
\begin{eqnarray}
\frac{d\sigma_{AA}}{dP_{J\perp}}&=&\int dxdyd\psi \frac{T_{AB}(x,y)}{2\pi}\int d\epsilon\nonumber\\
&&D(\epsilon,\omega_c(x,y,\psi;\hat{q}_0))\left.\frac{d\sigma_{pp}}{dP_{J\perp}'}\right|_{P_{J\perp}=P_{J\perp}'-\epsilon}
\end{eqnarray}
where $T_{AB}$ is the overlap normalization factor such that $\int dxdyd\psi \frac{T_{AB}(x,y)}{2\pi}=1$, with $P_{J\perp}'$ the partonic  jet transverse momentum, and $P_{J\perp}$ the observed(quenched) jet $p_T$.

The broadening effect of the medium also enters the Sudakov factor in an elegant form due to the fact that the vacuum radiations and medium effects contributes differently to the transverse momentum broadening in a well-separated regions of their phase space integral\cite{Mueller:2016gko,Mueller:2016xoc} given as follows:
\begin{equation}
S_{AA}(Q,b)=S_{pp}(Q,b)+\hat{q}_RL\frac{b^2}{4}
\end{equation}
where the contributions from the vacuum and medium were effectively factorized.

\begin{figure}
\begin{center}
\includegraphics[width=0.8\linewidth]{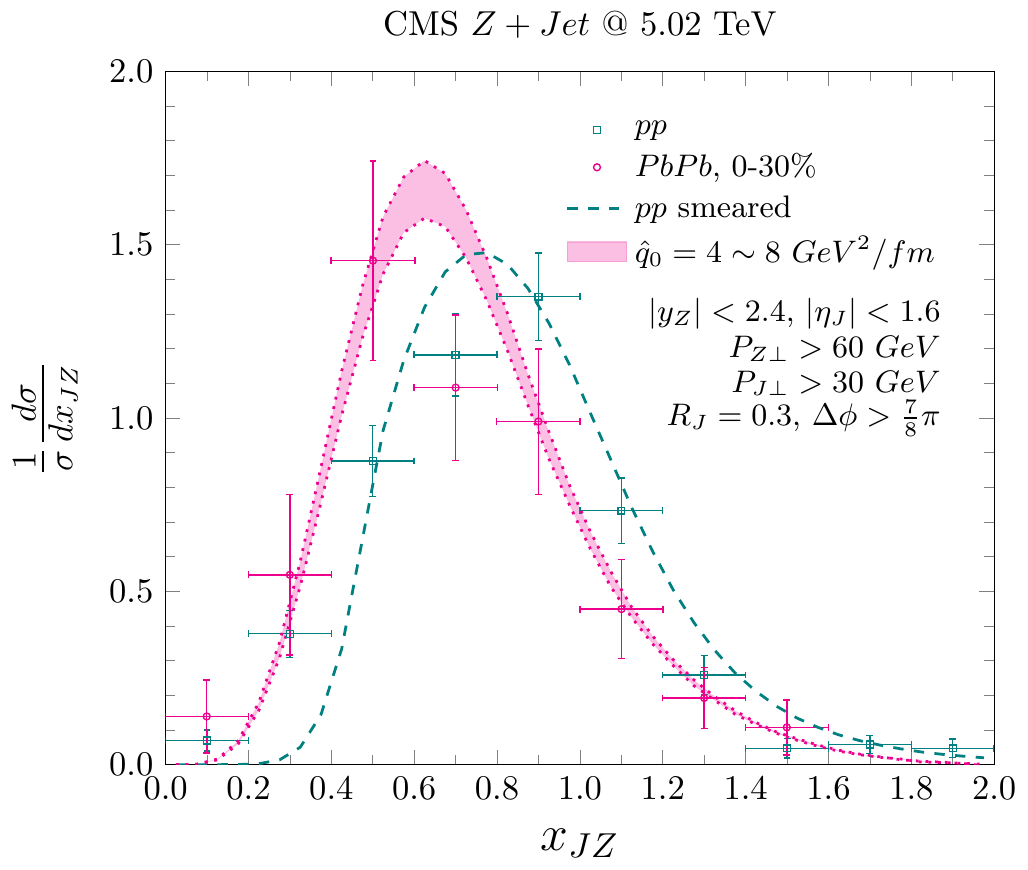}
\end{center}
\caption{Normalized smeared $x_{JZ}$ distribution for both $pp$ (dashed) and central $PbPb$ 0-30\% (dotted) data at 5.02 TeV in comparison with the CMS\cite{Sirunyan:2017jic} experimental data.}
\label{fig:5020ppaa}
\end{figure}
We then plotted the momentum imbalance ($x_{JZ}$) distribution in Fig. \ref{fig:5020ppaa} for both the $pp$ and central $PbPb$ at 0-30\% indicated by dashed and dotted lines respectively using the resummation formalism developed above. The results were compared to the CMS\cite{Sirunyan:2017jic} experimental data and both $pp$ and $AA$ distributions were normalized to unity.
Comparing nucleus-nucleus with nucleon-nucleon collisions, we see a shift of the distribution peak towards small $x_{JZ}$ both data and calculation indicating a clear sign of jet quenching due to the fact that jets loss energy when traversing through the QGP while the neutral boson remains the same, resulting a decrease in their ratio.
We found good agreement with the experimental data by setting the parameters with $\hat{q}_0=4\sim8~GeV^2/fm$ normalized to their central value from our previous studies.
By analysis, $D(\epsilon)$ is similar to a memoryless exponential(Poisson) distribution, and peaks around $\epsilon=5\sim10~GeV$ with the above setting, this agrees with the soft gluon approximation used by the BDMPS formalism. 

\begin{figure*}
\begin{center}
\includegraphics[width=0.3\linewidth]{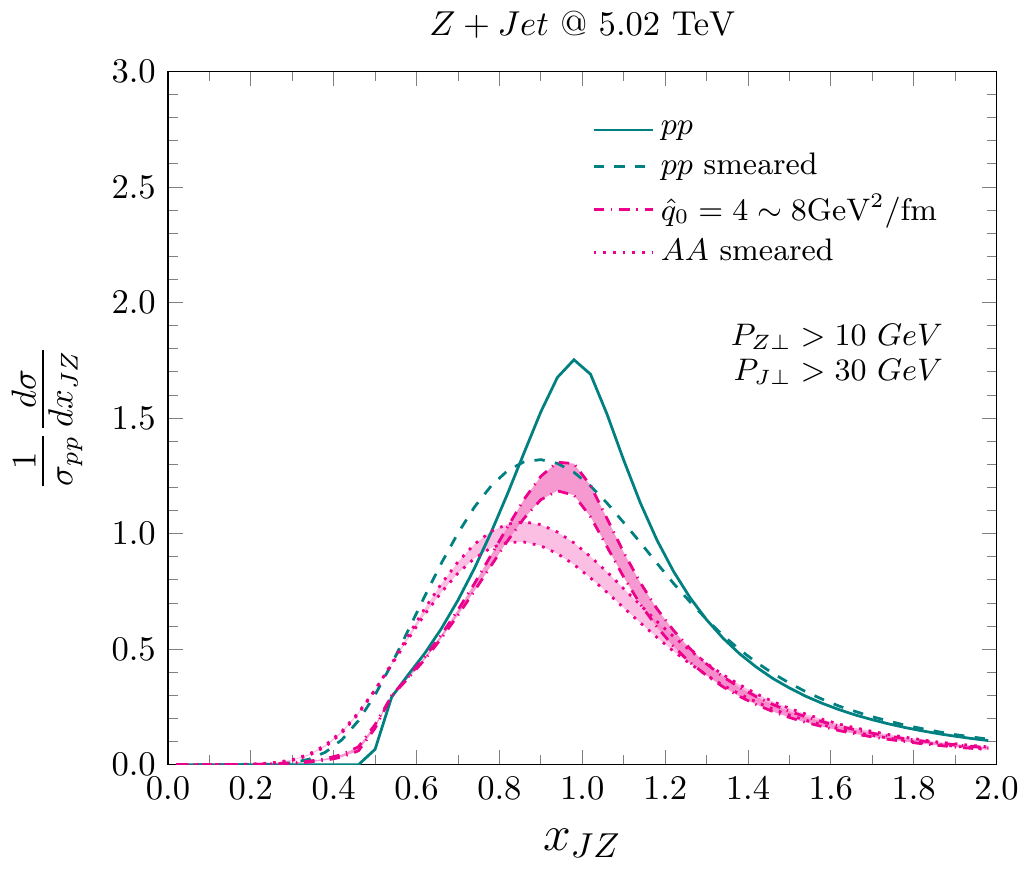}
\includegraphics[width=0.3\linewidth]{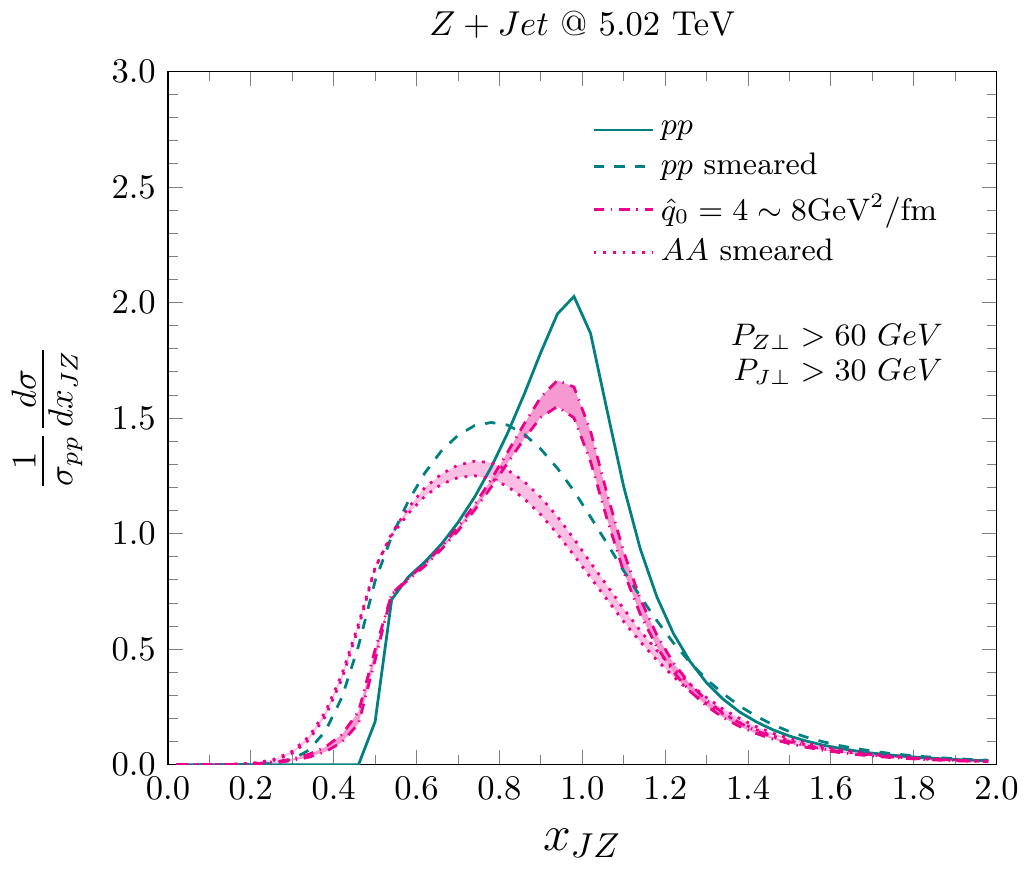}
\includegraphics[width=0.3\linewidth]{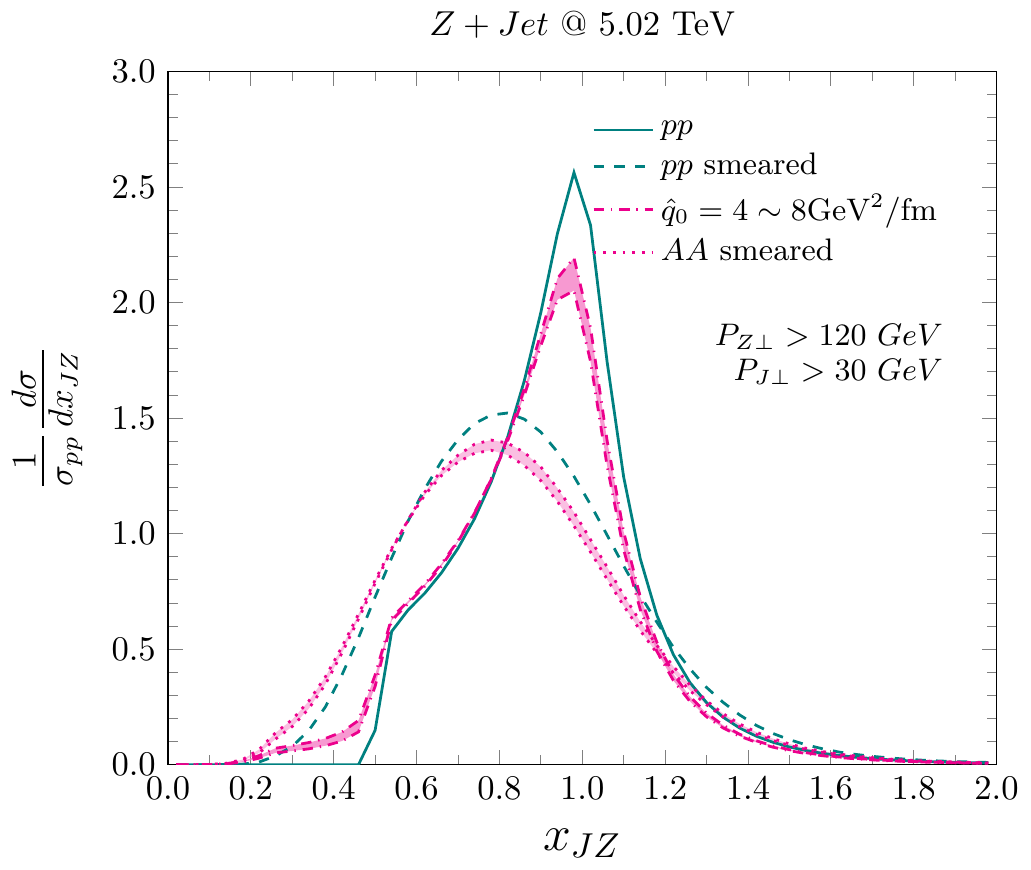}
\end{center}
\caption{A projection of the normalized $x_{JZ}$ distribution both smeared and unfolded for $pp$ and $PbPb$ collisions in the similar kinematic settings as in the $Z$+jet CMS\cite{Sirunyan:2017jic} experiment. Panels from left to right are $P_{Z\perp}>0$, 60 and 120 GeV respectively. $AA$ results are normalized to the $pp$ cross-section.}
\label{fig:5020aa}
\end{figure*}
We note that the calculation can be further improve by including the contribution of the $Y$-term, or shift $\phi_m$ towards $\pi$ to included higher order perturbative contribution that would otherwise raise the yield at small $x_J$ region. However, as demonstration, this study shows that resummation alone with certain kinematic selection could also provide good description to the experimental data, and we shall have a detail analysis on the effect of the additional $Y$-term or pQCD in our next study.

In Fig. \ref{fig:5020aa}, we provide a projection of the normalized unfolded $x_{JZ}$ distribution for both $pp$ and $PbPb$ collisions at 5.02 TeV in three $P_{Z\perp}$ ranges, which can help us narrow down the uncertainty and extract a more precise value of the transport coefficient. The sharp peak at $x_J=1$ corresponds directly to the back-to-back configuration due to the transverse momentum conservation, while the sharp drop at small $x_{JZ}$ corresponds to an implicit topological constrain. The small tail at the region $x_{JZ}>1$ as compared to the small $x_{JZ}$ shoulder suggest that the momentum of the boson is almost always larger than its associate jet, indicating that while jets undergo splittings, the boson remains inert. This shows that the boson is a good probe to calibrate the momentum of the jet, and higher order splittings will contribute more at small $x_{JZ}$ than large $x_{JZ}$. Because the $AA$ distribution are normalized to their corresponding $pp$ counterparts, we see a clear suppression of the overall distribution due to the loss of yield when jets loss enough energy to drop out of the kinematic cut, which also results in an overall shift to small $x_J$. Note that the right panel are $Z$ bosons with higher $P_{Z\perp}$ corresponds to high $P_{J\perp}$ jets by transverse momentum conservation. This means that quenching will result in less suppression and the $AA$ distribution will likely stay in shape as compared with panels on the left.

\begin{figure*}
\begin{center}
\includegraphics[width=0.3\linewidth]{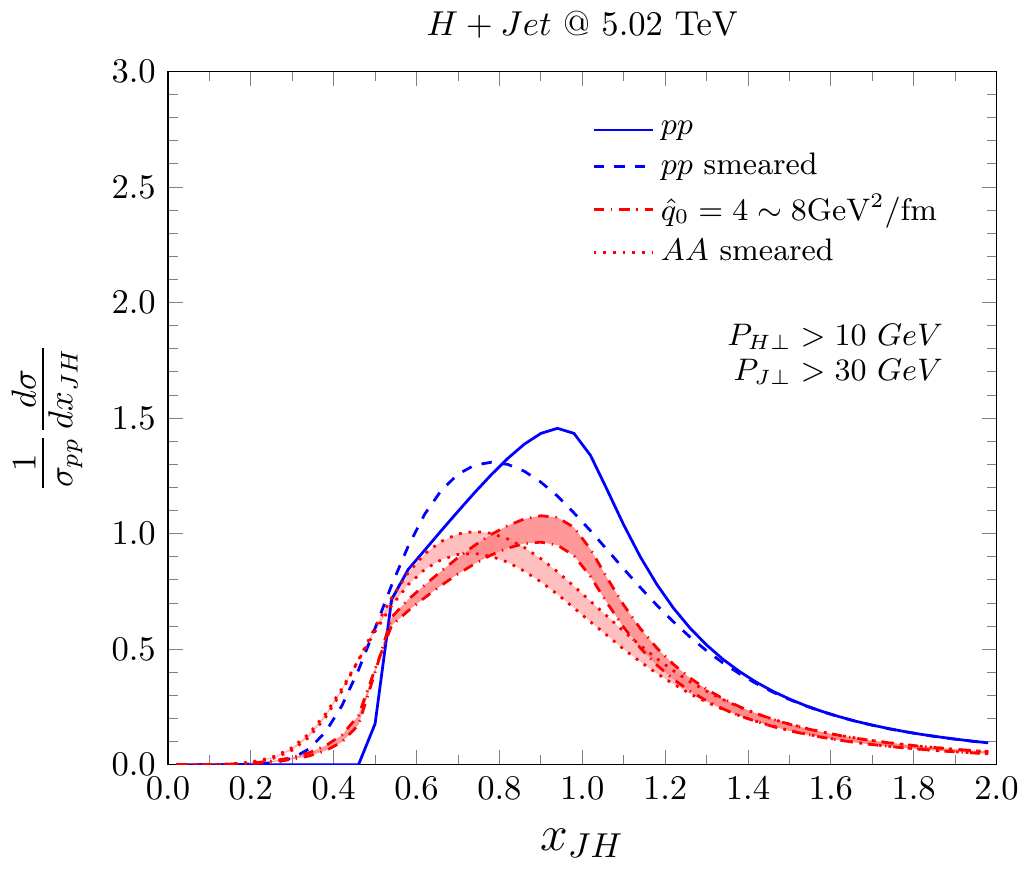}
\includegraphics[width=0.3\linewidth]{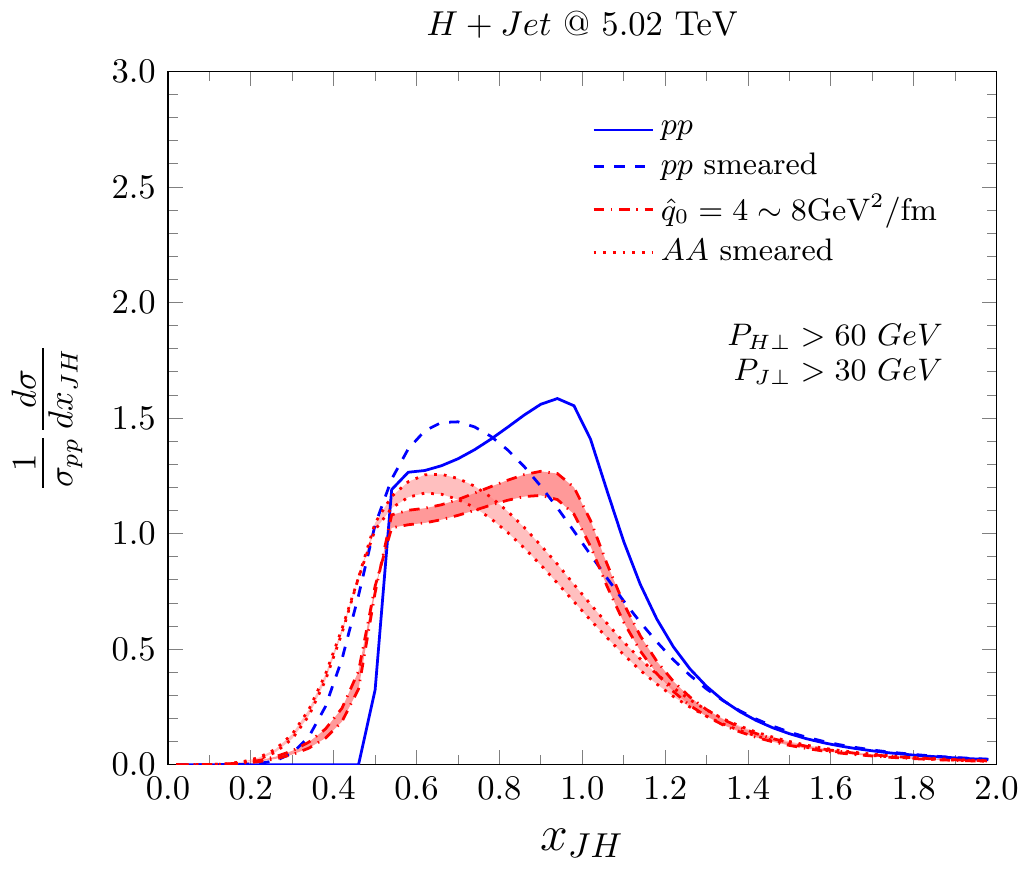}
\includegraphics[width=0.3\linewidth]{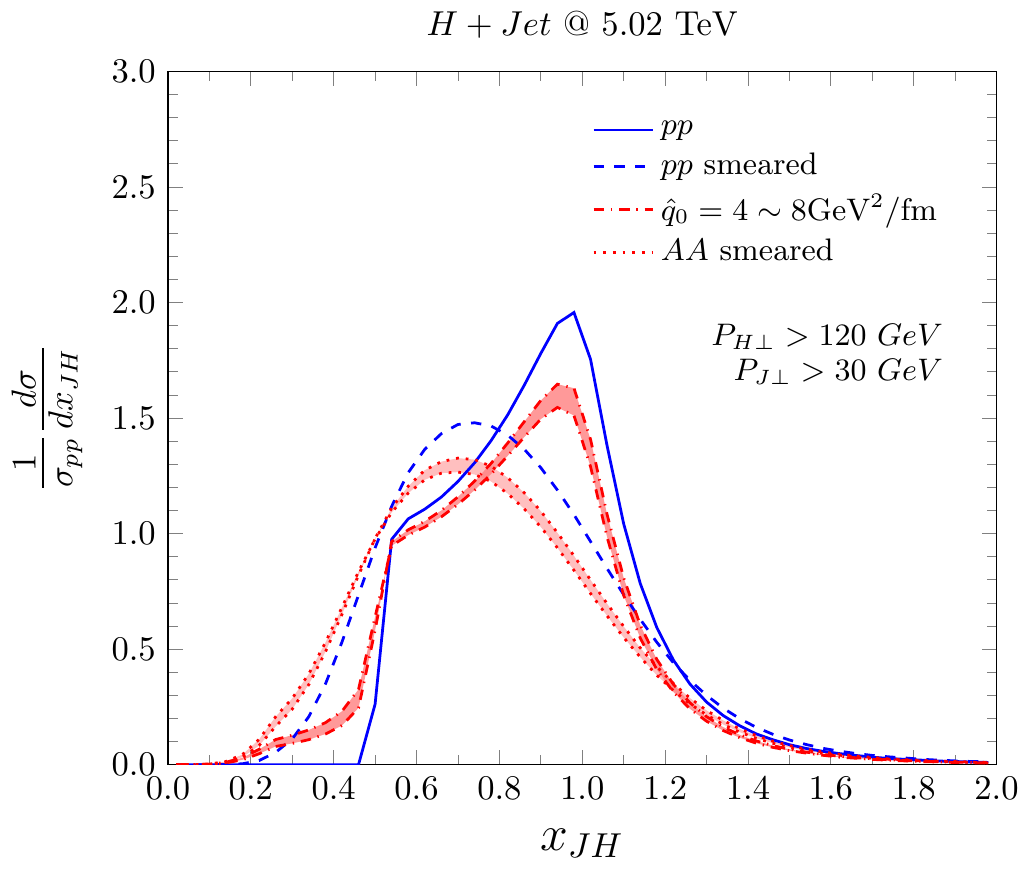}
\end{center}
\caption{A projection of the normalized $x_{JH}$ distribution both smeared and unfolded for $pp$ and $PbPb$ collisions in the same kinematic settings as in the $Z$+jet CMS\cite{Sirunyan:2017jic} experiment. Panels from left to right are $P_{H\perp}>0$, 60 and 120 GeV respectively. $AA$ results are normalized to the $pp$ cross-section.}
\label{fig:Hxj}
\end{figure*}
We then apply similar technique to the $H$+jet production assuming a future high luminosity upgrade to the LHC is available for statistical analysis. Using similar kinematic constrains and smearing parameters as $Z$-jet production, we see that $H$+jet production follows a very similar trend to $Z$+jet production shown in Fig. \ref{fig:Hxj}. Such that the $AA$(dashdotted) distribution as compared to $pp$(solid) will be suppresses due to quenching and the suppression will be significant at smaller $P_{H\perp}$. While the smeared distribution for $pp$(dashed) and $AA$(dotted) will give a smeared peak with a shift due to kinematic cuts. But because of the fact that Higgs have a stronger Sudakov effect previously shown, the large $q_\perp$ suppressed the $P_\perp$ of the associate jet, shifting the overall distribution towards smaller $x_J$, resulting in a broader peak at $x_J=1$.


\section{Conclusion}
\label{sec:conc}
Using the small $q_\perp$ resummation formalism, we have calculated the cross-section differential in $q_\perp$, $\Delta\phi$ and $x_J$ for both $Z$ and Higgs boson plus a Jet processes. We found that resummation alone is sufficient in the description of the $x_{JZ}$ distribution given the range of data selection, we then fixed the $pp$ baseline by comparing our theoretical results with the current $Z$+jet experimental data while fitting the smearing parameters. Then by the use of a hydro simulated profile and the BDMPS formalism, we plotted the quenched result with different $\hat{q}_0$ values in comparison with the $AA$ experimental data and found that it ranges around $\hat{q}_0=4\sim8~GeV^2/fm$. This agrees with our previous prediction of the $\gamma$-jet correlation and agrees also with other Monte-Carlo based energy-loss formalisms. We then provided a prediction of the unsmeared $x_{JZ}$ distribution and also $x_{JH}$ distributions for comparison to future experimental data. As a final remark, heavy boson tagged jets is an important high energy hard probe that could provide profound precision and great insights in the extraction of the transport coefficient of QGP.


\section*{Acknowledgements}
\label{sec:ack}
The authors would like to thank Bo-Wen Xiao and Guang-You Qin for the wonderful discussions and comments.
This work is supported by Natural Science Foundation of China (NSFC) under grant Nos. 11435004 and 11935007.
S.-Y. W. is supported by the Agence Nationale de la Recherche under the project ANR-16-CE31-0019-02.


\bibliographystyle{apsrev}
\bibliography{reference}

\end{document}